\documentclass[11pt]{article}
\usepackage{amsmath,amssymb}

\newcommand{\ft}[2]{{\textstyle\frac{#1}{#2}}}

\newsavebox{\uuunit}
\sbox{\uuunit}
    {\setlength{\unitlength}{0.825em}
     \begin{picture}(0.6,0.7)
        \thinlines
        \put(0,0){\line(1,0){0.5}}
        \put(0.15,0){\line(0,1){0.7}}
        \put(0.35,0){\line(0,1){0.8}}
       \multiput(0.3,0.8)(-0.04,-0.02){10}{\rule{0.5pt}{0.5pt}}
     \end {picture}}

\long\def\symbolfootnote[#1]#2{\begingroup%
\def\thefootnote{\fnsymbol{footnote}}\footnote[#1]{#2}\endgroup}
\numberwithin{equation}{section}

\begin{document}
\begin{titlepage}
\begin{center}
\hfill LMU-ASC 50/08 \\
\hfill ITP-UU-08/55 \\
\hfill SPIN-08/43 \\
\vskip 10mm

{\LARGE \textbf{The mixed black hole partition function\\[3mm] 
    for the STU model }}
\vskip 12mm

\textbf{G.L.~Cardoso$^{a}$, J.R.~David\symbolfootnote[1]{On lien from
  Harish-Chandra Research Institute, Allahabad, India}$^{b} $,
  B.~de~Wit$^c$ and  S.~Mahapatra$^{d}$} 

\vskip 6mm
$^a${\em Arnold Sommerfeld Center for Theoretical Physics\\
Department f\"ur Physik,
Ludwig-Maximilians-Universit\"at M\"unchen, Munich, Germany}\\
{\tt gabriel.cardoso@physik.uni-muenchen.de}\\[1mm]
$^b${\em Centre for High Energy Physics, Indian Institute of Science,
Bangalore 560 012, India} \\
{\tt justin@cts.iisc.ernet.in} \\[1mm]
$^c${\em Institute for Theoretical Physics} and {\em Spinoza
  Institute,\\ Utrecht University, Utrecht, Netherlands}\\
{\tt  B.deWit@uu.nl} \\[1mm]
$^d${\em 
Physics Department, Utkal University, 
Bhubaneswar 751 004, India}\\
{\tt swapna@iopb.res.in}\\[1mm]
\end{center}
\vskip .2in
\begin{center} {\bf ABSTRACT } \end{center}
\begin{quotation}\noindent
  We evaluate the mixed partition function for dyonic BPS black holes
  using the recently proposed degeneracy formula for the STU model.
  The result factorizes into the OSV mixed partition function times a
  proportionality factor. The latter is in agreement with the measure factor
  that was recently conjectured for a class of $N=2$ black holes that
  contains the STU model. 
\end{quotation}

\vfill
\end{titlepage}
\eject
\section{Introduction}
\label{sec:introduction}
\setcounter{equation}{0}
Some time ago it was conjectured that the partition function of
four-dimensional BPS black holes with $N=2$ supersymmetry, defined by
\begin{equation}
  \label{eq:mixed-partition} 
    Z(p,\phi) =
    \sum_{\{q\}} \; d(p,q) \, \mathrm{e}^{\pi \, q_I \phi^I } \,, 
\end{equation}
is related to the topological string partition function
\cite{Ooguri:2004zv}. The `mixed' partition function
\eqref{eq:mixed-partition} is based on an ensemble where the magnetic
charges $p$ and the electrostatic potentials $\phi$ are kept fixed.
With respect to the magnetic charges one is therefore dealing with a
micro-canonical ensemble, while the electric charges $q$ are replaced
by the continuous potentials $\phi$.  The $d(p,q)$ denote the
microscopic black hole degeneracies for given magnetic and electric
charges, $p^I$ and $q_I$, respectively.

The logarithm of the mixed partition function can be viewed
as a free energy function $\mathcal{F}_{\rm E}(p,\phi)$, 
\begin{equation}
  \label{eq:mixed-free-energy} 
    Z(p,\phi)  \sim  
    \mathrm{e}^{\pi\,\mathcal{F}_{\rm E}(p,\phi)}  \;,
\end{equation}
which can be identified with the one that exists in the context of the
field-theoretic description of BPS black holes. The latter has a
relation with the partition function $Z_\mathrm{top}(p,\phi)$ of the
topological string \cite{Bershadsky:1993cx}, which indicates that the
mixed partition sum \eqref{eq:mixed-partition} and the topological
string are related. In \cite{Ooguri:2004zv}, this relation was argued
to take the following form,
\begin{equation}
  \label{eq:FE=Ztop}
  \mathrm{e}^{\pi\,\mathcal{F}_{\rm E}(p,\phi)} = \vert
  Z_\mathrm{top}(p,\phi)\vert^2 \,. 
\end{equation}

In subsequent developments it was realized that, while $Z(p,\phi)$ is
invariant under certain imaginary shifts of the $\phi$ owing to the
quantized nature of the electric charges, this invariance is in
general not reflected in the free energy, so that one may have to
include an explicit sum over these shifts on the right-hand side of
\eqref{eq:mixed-free-energy}. Furthermore it turns out that
\eqref{eq:FE=Ztop} cannot be an exact relation, but must involve a
proportionality factor that plays the role of a measure in the inverse
Laplace transform that expresses the black hole degeneracies in terms
of the free energy.\footnote{
  There exist arguments of a more conceptual nature indicating that a
  modification of \eqref{eq:FE=Ztop} should be more drastic
  \cite{Cardoso:2008fr}. This issue is not directly relevant for the
  present paper, which mainly addresses \eqref{eq:mixed-free-energy}.
} 
In \cite{LopesCardoso:2006bg} it was shown how to determine this
measure from arguments based on duality in the context of the
semiclassical approximation of the inverse Laplace transform.
Independently, a direct evaluation of the mixed partition function
from specific microscopic degeneracy formulae for dyonic black holes
in $N=4$ supersymmetric CHL models \cite{Chaudhuri:1995fk} (carried
out in the context of an $N=2$ formalism) revealed the presence of a
measure factor \cite{Shih:2005he,LopesCardoso:2006bg}, which for large
charges was in agreement with the prediction of
\cite{LopesCardoso:2006bg} (see also \cite{de Wit:2007dn}).

These matters warrant further study in the context of $N=2$ black
holes, where not many degeneracy formulae are known. A proposal for
such a formula in the STU model \cite{Sen:1995ff,Gregori:1999ns},
which exhibits both exact S- and T-dualities, has been presented in
\cite{David:2007tq}. This proposal was considered in a recent paper
\cite{Cardoso:2008fr}, where a number of subtleties were noted (to
which we will turn in section \ref{sec:conclusion}), which, however,
stayed short of evaluating the measure factor. It is the purpose of
the present note to address this issue in more detail.

It is convenient to first discuss some common features
shared by the degeneracy formulae for the $N=4$ supersymmetric models
\cite{Dijkgraaf:1996it,Shih:2005uc,Jatkar:2005bh,David:2006ud} and
the STU model.\footnote{ 
  Formulae with torsion higher than one were constructed in 
  \cite{Banerjee:2008pv,Banerjee:2008pu,Dabholkar:2008zy}. 
} 
They all involve an integral over appropriate 3-cycles of the inverse
of an $\mathrm{Sp}(2,\mathbb{Z})$ automorphic form $\Phi_k
(\rho,\sigma,\upsilon)$ of weight $k$ (with suitable normalization), 
\begin{eqnarray}
   \label{eq:dphik}
   d_k(K,L,M) = I_k(K,L,M) = \oint \mathrm{d}
 \rho\,\mathrm{d}\sigma\,\mathrm{d}\upsilon \;  
 \frac{{\rm e}^{\mathrm{i} \pi [
     \rho\, K + \sigma\, L + (2 \upsilon -1)\,M]}}
 {\Phi_k(\rho,\sigma,\upsilon)} \;,
\end{eqnarray}
where $\rho,\sigma,\upsilon$ are the complex elements of the period
matrix of a genus-2 Riemann surface. The weight $k$ and the number $n$
of $N=2$ physical vector supermultiplets are related by
\begin{equation}
  \label{eq:k--n}
  n = 2(k+2) -1 \;.
\end{equation}
For the STU model we have $k = 0$ and $n=3$. For the $N=4$
supersymmetric models this relation changes into $n=2(k+2)+3$ to
account for the four gauge fields associated with the extra gravitini
supermultiplets; CHL black holes have been considered for
$k=1,2,4,6,10$. Upon including the $N=2$ graviphoton, there are thus
$n+1$ gauge fields, each associated with a magnetic and an electric
charge. These charges are denoted by $p^I$ and $q_I$, respectively,
where $I= 0,1,\ldots,n$.

The quantities $K,L,M$ take discrete values proportional to the charge
bilinears that transform as a triplet under an
$\mathrm{SL}(2,\mathbb{Z})$ factor of the duality group (or a subgroup
thereof) and are invariant under the remaining dualities. The explicit
transformations of $K,L,M$ are
\begin{eqnarray}
\label{eq:KLM-sl2}
  K &\to&  d^2 \,K +c^2 L +2\,cd\,M \,, \nonumber\\  
  L &\to&  a^2\,L +b^2\,K +2\,ab\,M \,, \nonumber\\
  M &\to& {}ac\, L +bd\,K +(ad+bc) M \,. 
\end{eqnarray}
For $\mathrm{SL}(2,\mathbb{Z})$ the parameters $a,b,c, d$ are
integer-valued parameters which satisfy $ad-bc =1$. For the $N=4$
models, these transformations constitute the S-duality group, which is
usually an arithmetic subgroup of $\mathrm{SL}(2,\mathbb{Z})$. For
the STU model this is either the S-, the T- or the U-duality group,
which equals the $\Gamma(2)$ subgroup of $\mathrm{SL}(2,\mathbb{Z})$.

The inverse of the automorphic form $\Phi_k$ takes the form of an
infinite Fourier sum with certain powers of
$\exp[\mathrm{i}\pi \rho]$, $\exp[\mathrm{i}\pi\sigma]$ and
$\exp[\mathrm{i}\pi \upsilon]$, and the 3-cycle is then defined by
choosing integration contours where the real parts of $\rho$, $\sigma$
and $\upsilon$ take the appropriate values to select the Fourier modes
in $1/\Phi_k$. Obviously the values taken by $(K,L,M)$ must be 
correlated with the possible Fourier modes. The leading behaviour of
the dyonic degeneracy is associated with the rational quadratic
divisor ${\cal D} = \upsilon + \rho \sigma - \upsilon^2=0$ of
$\Phi_{k}$, near which $1/\Phi_k$ takes the form, 
\begin{eqnarray}
   \label{eq:Phik}
   \frac{1}{\Phi_k (\rho,\sigma,\upsilon)}  \approx
   \frac{1}{4\,\pi^2}\, 
   \frac{1}{\mathcal{D}^2} \; \frac1{\Delta_k(\rho,\sigma,\upsilon)} 
   +\mathcal{O}(\mathcal{D}^0) \;,\qquad \Delta_k = \frac{
   f^{(k)}(\gamma')\,f^{(k)}(\sigma')}{\sigma^{k+2}} \,,
\end{eqnarray}
where
\begin{eqnarray}
   \label{eq:gsprime}
   \gamma' = \frac{ \rho \sigma - \upsilon^2}{\sigma} \;,\qquad 
 \sigma' = \frac{\rho \sigma - (\upsilon-1)^2}{\sigma} \;.
\end{eqnarray} 
Here $f^{(k)}$ is a known modular form associated with
$\mathrm{SL}(2,\mathbb{Z})$ or its appropriate subgroup. For the STU
model we have 
$f^{(0)}(\gamma^\prime)=\vartheta_2^{\,4}(\gamma^\prime)$.

The choice of the divisor $\mathcal{D}$ strongly restricts possible
redefinitions of the complex variables $\rho,\sigma,\upsilon$.  Both
the exponential factor in \eqref{eq:dphik} and the divisor are
invariant under the dualities corresponding to \eqref{eq:KLM-sl2}, which
implies,
\begin{eqnarray}
\label{eq:modparatrans}
    \rho &\to& a^2\,\rho +b^2\,\sigma -2\,ab\,\upsilon +ab\,, \nonumber\\
  \sigma &\to& c^2 \rho + d^2 \,\sigma -2\,cd\,\upsilon+cd\,, \nonumber\\  
  \upsilon &\to& {}-ac\, \rho -bd\,\sigma +(ad+bc) \upsilon-bc \,.
\end{eqnarray}
These transformations belong to the modular group
$\mathrm{Sp}(2,\mathbb{Z})$ associated with $\Phi_k$. The
inhomogeneous terms in \eqref{eq:modparatrans} contribute only to the
real part of $\rho,\sigma,\upsilon$, and they have some bearing on the
periodicity intervals for the real values of $\rho,\sigma,\upsilon$.

For the $\Gamma(2)$ subgroup of $\mathrm{SL}(2,\mathbb{Z})$, which is
relevant for the STU model, we have $a,d=1+2\,\mathbb{Z}$ and
$b,c=2\,\mathbb{Z}$, so that the real shifts induced in
$\rho,\sigma,\upsilon$ are multiples of 2. This is consistent with
the fact that $1/\Phi_0$ has a Fourier decomposition in terms of
powers of $\exp[\mathrm{i}\pi\rho]$, $\exp[\mathrm{i}\pi\sigma]$ and
$\exp[2\mathrm{i}\pi\upsilon]$, which implies that $K$, $L$ and $M$
must take integer values in order to find  non-zero values for
\eqref{eq:dphik}. Hence, the 3-cycle can be parametrized by, 
\begin{equation}
  \label{eq:cycles-STU}
  0\leq\mathrm{Re}\,\sigma < 2\,,\qquad 0\leq\mathrm{Re}\,\rho < 2\,,
  \qquad 0\leq\mathrm{Re}\,\upsilon < 1\,. 
\end{equation}
The lattice of the charges $p^I$ and $q_I$ will be discussed in the
next section.

The proposal of \cite{David:2007tq} for the dyon degeneracy in the STU
model involves three integrals of the type \eqref{eq:Phik}, and reads,
\begin{equation}
  \label{eq:deg-STU}
  d_\mathrm{STU}(p,q) = I_0(K_s,L_s,M_s)\;I_0(K_t,L_t,M_t)\;
  I_0(K_u,L_u,M_u)\;, 
\end{equation}
which is manifestly invariant under triality (related to interchanging
the $s$, $t$ and $u$ labels), where the triplets of charge bilinears,
$(K_s,L_s,M_s)$, $(K_t,L_t,M_t)$ and $(K_u,L_u,M_u)$, transform as
vectors under S-, T- and U-duality, respectively.

An asymptotic evaluation of the integral (\ref{eq:dphik}) can be done
in the limit where $K L-M^2\gg 1$ and $K+L$ is large and negative.
Furthermore one assumes that $\vert K\vert$ is sufficiently small as
compared to $\sqrt{KL -M^2}$. In this way one can recover
non-perturbative string corrections, as was stressed in
\cite{Cardoso:2004xf}.  The evaluation of the integral
\eqref{eq:dphik} proceeds by first evaluating the contour integral for
$\upsilon$ around either one of the zeros $\upsilon_{\pm}= \frac12 \pm
\frac12 \sqrt{1 + 4 \rho \, \sigma}$ of $\Phi_k$ on the divisor ${\cal
  D} = 0$.  Subsequently, the two remaining integrals over $\rho$ and
$\sigma$ are evaluated in saddle-point approximation. The saddle-point
values of $\rho, \sigma$, and hence of $\upsilon_{\pm}$ are expressed
in terms of $\sigma'$ and $\gamma'$ in a way that is independent of
the choice of the pole position $\upsilon_\pm$. As it turns out,
$\sigma'$ and $\gamma'$ can be identified with the complex modulus $S$
in a field-theoretic description, according to $\gamma' = {\rm i}S$
and $\sigma' ={\rm i}{\bar S}$ \cite{Cardoso:2004xf}. In that case the
saddle-point values of $\rho$, $\sigma$ and $\upsilon$ can be
parametrized by
\begin{equation}
  \label{eq:rhosigups-asympt}
  \rho = \frac{{\rm i } |S|^2}{S + {\bar S}} \;,\qquad 
  \sigma = \frac{\rm i}{S + \bar S} \;,\qquad  \upsilon = 
  \frac{S}{S + \bar S} \;. 
\end{equation}
These values describe the unique solution to the saddle-point
equations for which $d_k(K,L,M)$ takes a real value. The resulting
expression for $\ln d_k (K,L,M)$ equals
\begin{equation}
  \label{eq:value-dk-saddle}
  \ln d_k(K,L,M) =  \pi \left[ - 
  \frac{L - {\rm i} M (S - \bar S)
    + K |S|^2}{S + {\bar S}} - \frac1{\pi} \ln \Delta_k (S, \bar
  S)\right]  \;,
\end{equation}
where the right-hand side is evaluated at a stationary point, so that
$S$ (and therefore $\rho,\sigma,\upsilon$) will be determined in terms
of $K,L,M$. In the limits specified earlier it turns out that $S$
takes a finite value. The result then coincides precisely with the
results obtained in the field-theoretic description
\cite{Cardoso:2004xf,Jatkar:2005bh,LopesCardoso:2006bg}, and it holds
up to an additive constant and up to terms that are suppressed by
inverse powers of the charges.  Substituting the value for $S$ and
working to first order in $\Delta_k$ gives
\begin{equation}
  \label{eq:approx-entropy}
    \ln d_k(K,L,M) \approx  \pi \sqrt{K\,L-M^2} - \ln \Delta_k (S,\bar S) \;.
\end{equation}

This paper is organized as follows. In section
\ref{sec:prot-mixed-part} we consider the typical calculation of a
mixed partition function, which will be used in the evaluation of the
full mixed partition function for the STU model in section
\ref{sec:mixed-part-funct-STU}. In section \ref{sec:conclusion} we
present our conclusions.

\section{Prototype evaluation of the mixed partition function}
\label{sec:prot-mixed-part}
\setcounter{equation}{0}

In the following, we compute the mixed partition function associated
with $d_k(p,q)$ expressed by the integral \eqref{eq:dphik}. For
definiteness we will be more specific here and consider the STU model,
where $k=0$. We follow the same strategy as in
\cite{Shih:2005he,LopesCardoso:2006bg} where the $N=4$ supersymmetric
models were considered.

As indicated in \eqref{eq:deg-STU}, the degeneracies for the STU model
factorize into three integrals of the type \eqref{eq:dphik}, which are
related by triality.  Here we will first evaluate the mixed partition
function as if there is only one such integral,
\begin{equation}
  \label{eq:Z-s}
  Z_s(p,\phi) = \sum_q \,d_0(K_s,L_s,M_s)\, \mathrm{e}^{\tfrac13 \pi\,
  q_I\phi^I} \;.
\end{equation}
The reason for the factor $\tfrac13$ in the exponent will become clear
in the next section, where we will use the result of this calculation
to obtain the full expression for the STU model based on
\eqref{eq:deg-STU}.

According to \cite{David:2007tq}, the fact that we have three such
integrals implies that the quantities $(K_s,L_s,M_s)$ must be equal to
one-third of the charge bilinears $(\langle P,P\rangle_s,\langle
Q,Q\rangle_s,\langle P,Q\rangle_s)$ that were used in the supergravity
formulation (we use the notation of \cite{Cardoso:2008fr}). Hence 
we have (note that $I=0,1,2,3$),
\begin{eqnarray}
  \label{eq:KLM-pq}
  3\,K_s &=&{}\langle P,P\rangle_s ~=~ - 2(p^0 q_1 +p^2p^3 ) \;, 
  \nonumber\\
  3\, L_s&=&{}\langle Q,Q\rangle_s ~=~  2(q_0 p^1 - q_2 q_3) \;,
  \nonumber\\
  3\, M_s &=& \langle P,Q\rangle_s ~=~ 
  q_0 p^0 - q_1 p^1 + q_2 p^2 + q_3p^3 \;.
\end{eqnarray}
It is clear that the charges $p^I$ and $q_I$ cannot be integer-valued
in this case, in view of the fact that the three quantities $K_s$,
$L_s$ and $M_s$ must cover the same set of integer values. Combining
various arguments presented in the previous section, we therefore
conclude that the charges $p^I$ and $q_I$ take the following values,
\begin{equation}
  \label{eq:rescale-pq}
    p^{1,2,3}, q_0 \in \lambda^{-1}\,\mathbb{Z} \,,\qquad 
    p^0, q_{1,2,3}\in  \lambda\,\mathbb{Z}\,,
\end{equation}
where $\lambda= \sqrt{2}$ or $\tfrac12\sqrt{2}$, which is consistent
with triality. The reader can easily verify that, based on the
possible values of $p^I$ and $q_I$, $K_s$, $L_s$ and $M_s$ cover the
full range of integers (as well as the rational numbers
$\mathbb{Z}\pm\tfrac13$ for which $I_0$ will vanish). In the effective
action the two values of $\lambda$ are simply related by a uniform
electric/magnetic duality transformation under which all $p^I$ and
$q_I$ are interchanged. The above assignment is consistent with string
theory where the STU model is described in terms of a freely acting
$\mathbb{Z}_2\times\mathbb{Z}_2$ orbifold of type-IIB string theory
compactified on $T^4\times S^1\times \tilde S^1$. Here the choice of
$\lambda$ is related to the identification of the $p^I$ and $q_I$ with
the momenta and winding numbers associated with the two circles, $S^1$
and $\tilde S^1$. We will not make a choice for $\lambda$ in what
follows in order to make the effect of the charge basis explicit in
the calculation.

We now proceed and follow the derivation as presented in
\cite{LopesCardoso:2006bg}, recalling that while the charges $p^I$ and
$q_I$ take the values given in \eqref{eq:rescale-pq}, \eqref{eq:dphik}
will only be nonvanishing for $K_s,L_s,M_s \in \mathbb{Z}$.  The
integration contours are chosen according to \eqref{eq:cycles-STU},
and $1/\Phi_0$ can be expanded in terms of Fourier coefficients $\exp
\mathrm{i}\pi[m\rho+n\sigma+2p\upsilon]$ with integers $m,n,p$.

We begin by summing over $q_0$ and $q_1$, replacing the sums over
$q_0$ and $q_1$ in \eqref{eq:Z-s} by sums over the charges $L_s$ and
$K_s$, related by the identities,
\begin{equation}
  \label{eq:Q-P-ids}
  q_0 = \frac1{2\,p^1} \left(3 L_s+ 2\,q_2q_3\right) \,,\qquad    
  q_1 = - \frac1{2\,p^0} \left(3 K_s +2\, p^2p^3\right) \,.
\end{equation}
We will be assuming that both $p^0$ and $p^1$ are non-vanishing and
positive (the latter is only a matter of convenience). In doing so, we
need to ensure that, when performing the sums over $L_s$ and $K_s$, we
only keep those contributions that lead to integer-valued charges of
$\lambda q_0$ and $q_1/\lambda$. This projection onto integer values
can be implemented by inserting the series $N^{-1}\sum_{l=0}^{N-1}
\;\exp[2\pi\mathrm{i} \,l\, P/N]$,
where $P$ and $N$ are integers,\footnote{ 
  We assume that $N\geq1$. Note that this formula remains correct when
  $P$ and $N$ have a common divisor. } 
which projects onto all integer values for $P/N$.  The use of this
formula leads to the following expression,
\begin{eqnarray}
\label{eq:zQP}
  Z_s(p,\phi) &=& \frac{1}{4 p^0 p^1 }\sum_{
    \begin{array}{l} \scriptstyle
      \phi^0  \rightarrow\phi^0+ 6\mathrm{i}\, l^0\lambda \\[-.2mm]
      \scriptstyle 
      \phi^1 \rightarrow\phi^1+ 6\mathrm{i} \,l^1/\lambda \end{array} }
  \sum_{L_s,K_s,q_2,q_3}  \, d_0(L_s,K_s,M_s)  \\
  &&\times \exp \left[\frac{\pi \phi^0}{6p^1}(3L_s +2q_2 q_3)  -
      \frac{ \pi \phi^1}{6p^0} (3K_s +2p^2 p^3) + \frac{\pi}{3}\,
      (q_2 \phi^2+q_3\phi^3)    \right] \;, \nonumber
\end{eqnarray}
with $M_s$ given by
\begin{equation}
  \label{eq:Ms-q}  
  M_s = \frac{p^0}{6\,p^1}(3L_s + 2\,q_2q_3)+
  \frac{p^1}{6\,p^0}(3K_s + 2 p^2 p^3) + \frac13(q_2 p^2 + q_3 p^3) \,.
\end{equation}
In (\ref{eq:zQP}) the summation over imaginary shifts of $\phi^0$ and
$\phi^1$ is implemented by first replacing $\phi^{0}
\rightarrow\phi^{0}+ 6\,\mathrm{i} l^{0}\lambda$ and $\phi^{1}
\rightarrow\phi^{1}+ 6\,\mathrm{i} l^{1}/\lambda$ in each summand, and
subsequently summing over the integers $l^0 = 0,\ldots, 2
p^1\lambda^{-1}-1$ and $l^1 = 0,\ldots,2p^0\lambda-1$. The sums over
$l^{0,1}$ enforce that only those summands for which $(3\,L_s + 2\,
q_2 q_3)\lambda/2p^1$ and $(3\,K_s + 2\,p^2p^3)/2p^0\lambda$ are
integers, give a non-vanishing contribution to $Z_s(p,\phi)$.

Next, consider summing over $L_s$ without any restriction. Expanding
$1/\Phi_k$ in Fourier modes,
\begin{equation}
  \label{eq:F-sum-sigma}
  \frac{1}{\Phi_0 (\rho, \sigma,\upsilon)} 
  = \sum_n {\rm e}^{{\rm i}\pi\, n \sigma} C_n (\rho,\upsilon) \;,
\end{equation}
results in the double sum
\begin{equation}
  \label{eq:double-sum-L}
  \sum_{L_s,n} {\rm e}^{{\rm i}\pi\, [L_s(\sigma -\sigma(\upsilon) ) +
  n\,\sigma]} \,  C_n (\rho,\upsilon) \;,
\end{equation}
where we introduced
\begin{equation}
\label{eq:sigma(v)}
\sigma(\upsilon)= -\frac{\phi^0}{2\mathrm{i} p^1} - (2\upsilon-1)
   \frac{p^0}{2p^1}\,.
\end{equation}
Subsequently, consider performing the contour integral of
\eqref{eq:double-sum-L} over $\sigma$.  This selects the Fourier mode
$n = - L_s$, so that we obtain, 
\begin{equation}
  \label{eq:F-sigma}
  2\,\sum_{L_s} {\rm e}^{{\rm i}\pi L_s\sigma(v) } \,
  C_{L_s} (\rho, v) = \frac{2}{\Phi_0 (\rho,\sigma(\upsilon),\upsilon)} 
  \;.
\end{equation}
Next, summing over $K_s$ without any restriction and using
analogous steps as described above, yields 
\begin{equation}
  \label{eq:F-sigma-rho}
  \frac{4}{\Phi_0 (\rho(\upsilon), \sigma(\upsilon), \upsilon)} \;,
\end{equation}
where
\begin{equation}
  \label{eq:rho(v)}
  \rho(\upsilon)= \frac{\phi^1}{2\mathrm{i} p^0} - (2\upsilon-1)
   \frac{p^1}{2p^0}\,.
\end{equation}
Hence, after summing over $L_s$ and $K_s$ and performing two of the three
contour integrals, we obtain
\begin{eqnarray}
  \label{eq:mpfpk}
   && Z_s(p,\phi)= \frac{1}{p^0  p^1}
    \sum_{ \begin{array}{l} \scriptstyle
      \phi^0  \rightarrow\phi^0+ 6\mathrm{i}\, l^0\lambda\\[-.2mm] 
      \scriptstyle 
      \phi^1 \rightarrow\phi^1+ 6\mathrm{i} \,l^1/\lambda \end{array} }
  \sum_{q_2,q_3}\; 
  \oint\,\mathrm{d} \upsilon\;\frac{1}
  {\Phi_{0}(\rho(\upsilon),\sigma(\upsilon),\upsilon)} \nonumber\\
  &&\!
  \times \exp\left(- \tfrac13\mathrm{i}\pi  
    \left[2\, \sigma(\upsilon) \, q_2 q_3
      +2\,\rho(\upsilon)\, p^2 p^3 + \mathrm{i} q_2(\phi^2
    +\mathrm{i} (2\upsilon-1) p^2) + \mathrm{i} q_3(\phi^3
    +\mathrm{i} (2\upsilon-1) p^3)  
    \right] \right) \,.\nonumber\\
  &&{~} 
\end{eqnarray}
The integrand is manifestly invariant under the shifts $\phi^0\to
\phi^0+ 3\,\mathrm{i} p^1$, $\phi^1\to \phi^1+12\,\mathrm{i} p^0$ (or
$\phi^0\to \phi^0+ 12\,\mathrm{i} p^1$, $\phi^1\to
\phi^1+3\,\mathrm{i} p^0$, depending on the value of $\lambda$) and
$\phi^{2,3}\to \phi^{2,3} + 6\mathrm{i}\lambda^{-1}$, so that the
explicit sum over shifts with $l^0 = 0, \ldots, 2p^1\lambda^{-1}-1$
and $l^1 = 0,\ldots, 2p^0\lambda-1$ ensures that the partition
function (\ref{eq:mpfpk}) is invariant under any shifts of
$\phi^{1,2,3}$ that are multiples of $6\mathrm{i}\lambda^{-1}$ and of
$\phi^0$ that are multiples of $6\mathrm{i}\lambda$. Note that we are
overcounting in this way, because the full range of the explicit sum
over shifts of either $\phi^0$ or $\phi^1$ is not required in view of
the explicit invariance of the integrand. In this particular case this
will lead to an irrelevant multiplicative factor 4. In practice we
will impose an infinite sum over shifts for all the fields $\phi$,
while modding out the shifts that correspond already to an invariance
in the final result.  In this way we respect the symmetry of the
initial expression \eqref{eq:mixed-partition}.

Subsequently we perform a formal Poisson resummation over the charges
$q_2$ and $q_3$, and obtain\footnote{ 
  The resummation involves a (divergent) gaussian integral, which can
  be evaluated upon performing an analytic continuation of the
  integration variables. We assume that this continuation leads to an
  overall factor $-\mathrm{i}$. Observe that we are interested in the
  result for imaginary values of $\sigma(\upsilon)$, as is explained
  below. } 
\begin{eqnarray}
\label{eq:partpoisson}
   Z_s(p,\phi) &=& {}-\frac{3\,\mathrm{i}} {\lambda^2 p^0 p^1} \,  \,
     \sum_{\mathrm{shifts}}  
     \oint \,\mathrm{d}\upsilon \; \frac{1}
       {\sigma (\upsilon) \; \Phi_0 (\rho(\upsilon), \sigma
       (\upsilon), \upsilon)} \nonumber\\
     && \times \exp \left( - \tfrac13\mathrm{i}\,\pi 
       \left[2\,p^2p^3\,\rho(\upsilon)  
         + \frac{ ( \phi^2 +\mathrm{i} (2\upsilon-1)p^2) \, 
           (\phi^3 +\mathrm{i}(2\upsilon-1)p^3)}
         {2 \sigma (\upsilon)} \right]\right) \;, \nonumber\\
     &&{~} 
\end{eqnarray}
where the sum over shifts now also includes an infinite sum over
multiple shifts 
$\phi^{2,3}\rightarrow\phi^{2,3}+6\mathrm{i}\lambda^{-1}$, which are
induced by the Poisson summation. Note that the invariance over the
shifts $\phi^0\to \phi^0+ 3\,\mathrm{i} p^1$ (or $\phi^0\to \phi^0+
12\,\mathrm{i} p^1$) is no longer manifest after the resummation.
 
Now we perform the contour integral over $\upsilon$. This integration
picks up the contributions from the residues at the various poles of
the integrand. We assume that the leading contribution to this sum of
residues stems from the rational quadratic divisor ${\cal D} =
\upsilon + \rho \sigma - \upsilon^2=0$ of $\Phi_{0}$. Other poles of
the integrand in (\ref{eq:partpoisson}) are expected to give rise to
exponentially suppressed contributions in the limit that the charges
are large. Inserting $\rho(\upsilon)$ and $\sigma(\upsilon)$ into
${\cal D}$ yields
\begin{equation}
\label{eq:D-dyonic}
  {\cal D}= 2(\upsilon-\upsilon_*) \,\frac{\phi^0 p^1 - \phi^1 p^0}
  {4\mathrm{i} p^0p^1}\,,
\end{equation}
with $\upsilon_*$ given by 
\begin{equation}
\label{eq:v-star}
2\upsilon_* = 1-\mathrm{i}\, 
\frac{\phi^0\phi^1+p^1p^0}{\phi^0p^1-\phi^1p^0}\,.
\end{equation}
The corresponding values of $\rho_*=\rho(\upsilon_*)$ and
$\sigma_*=\sigma(\upsilon_*)$ take the following form, 
\begin{equation}
  \label{eq:sigma-rho*}
  \sigma_* = \frac{\mathrm{i}}{2}\,
  \frac{(\phi^0)^2+(p^0)^2}{\phi^0p^1-\phi^1p^0}\,, \qquad
  \rho_* = \frac{\mathrm{i}}{2}\,
  \frac{(\phi^1)^2+(p^1)^2}{\phi^0p^1-\phi^1p^0}\,. 
\end{equation}

We observe that ${\cal D}$ has only a simple zero. Using
\eqref{eq:Phik} we can perform the contour integral over $\upsilon$,
which yields,
\begin{equation} 
  \label{eq:zres}
  \begin{split}
    Z_s(p,\phi) &={}- \frac{6\,p^0 p^1}{\pi\lambda^2} 
      \sum_{\mathrm{shifts}}\; \frac{1}{(\phi^0p^1-\phi^1p^0)^2}
      \\[2mm] 
  &\times \frac{\mathrm{d}}{\mathrm{d}\upsilon}\left[
  \frac{
    \exp \left( - \tfrac13\mathrm{i}\pi \left[2 p^2 p^3 \,
      \rho(\upsilon) + \frac{ ( \phi^2 +\mathrm{i} 
          (2\upsilon-1)p^2) \,( \phi^3 +\mathrm{i}
          (2\upsilon-1)p^3)}{2 \sigma (\upsilon)}  \right] \right)}
          {\sigma(\upsilon)\,
      \Delta_0(\rho(\upsilon),\sigma(\upsilon),\upsilon)}
      \right]_{\upsilon = \upsilon_*}\,.
 \end{split}
\end{equation}
Evaluating this expression leads to (we refer to
\cite{LopesCardoso:2006bg} for additional details), 
\begin{eqnarray} 
  \label{eq:zres2}
    Z_s(p,\phi) &=& \frac2{\lambda^2}
    \sum_{\mathrm{shifts}}\; 
    \left[\frac{ \pi(p^2\phi^0 - p^0 \phi^2)
    (p^3\phi^0 - p^0 \phi^3) -6\mathrm{i}p^0\sigma_*\left(p^0 +
    p^1\sigma_*
    \frac{\mathrm{d}\ln[\sigma^2\Delta_0] }
    {\mathrm{d} \upsilon}\Big\vert_{*}\right)}  
  {\pi\,((\phi^0)^2 + (p^0)^2)(\phi^0p^1-\phi^1p^0)}\right]   \nonumber
    \\[.5ex] 
    &&\hspace{1.2cm} \times\,  \exp\left[\ft13\pi
    \mathcal{F}_0(p,\phi) - \ln[\sigma_*^2\;
        \Delta_0(\rho_*,\sigma_*,\upsilon_*) ] \right]\;, 
\end{eqnarray}
where 
\begin{eqnarray}
  \label{eq:mixed-free-energy-stu}
  \mathcal{F}_0(p,\phi) &=& \frac{-1}{(\phi^0)^2+(p^0)^2}\, 
        \Big[\phi^0(p^1\phi^2\phi^3 + p^2\phi^3\phi^1 +
        p^3\phi^1\phi^2)  \nonumber \\
        &&{}\hspace{2.4cm} +
        p^0(\phi^1p^2p^3 + \phi^2p^3p^1 + \phi^3p^1p^2)  
        -p^0 \phi^1\phi^2\phi^3 - \phi^0 p^1p^2p^3\Big]\,,  \nonumber\\ 
        &&{~} 
\end{eqnarray}
which is manifestly invariant under triality.

We close this section by indicating the relationship with various
quantities that appear in the macroscopic description of the STU
model. First we define $Y^I$ by
\begin{equation}
  \label{eq:def-Y}
  Y^I= \tfrac12(\phi^I+ \mathrm{i}p^I)\,,
\end{equation}
and we introduce the ratios $\mathrm{i}S= Y^1/Y^0$, $\mathrm{i}T=
Y^2/Y^0$ and $\mathrm{i}U= Y^3/Y^0$. It then follows straightforwardly
that
\begin{equation}
  \label{eq:rosigups-star}
  \rho_*=   \frac{\mathrm{i} \vert S\vert^2}{S+\bar S} \,,\quad 
  \sigma_* = \frac{\mathrm{i}}{S+\bar S}\,,\quad 
  \upsilon_*= \frac{S}{S+\bar S}\,,
\end{equation}
which coincides with \eqref{eq:rhosigups-asympt}. In this parametrization
it is easy to show that $\mathrm{Im}(\rho_*)\,\mathrm{Im}(\sigma_*) -
(\mathrm{Im}(\upsilon_*))^2 =\tfrac14$, so that the point
$(\rho_*,\sigma_*,\upsilon_*)$ is located on the Siegel upper-half
plane. Furthermore, we find that
\begin{equation}
  \label{eq:Delta-divisor}
  \omega(p^0,p^1,\phi^0,\phi^1)\equiv 
  \sigma_*^2\, \Delta_0(\rho_*,\sigma_*,\upsilon_*) =
  f^{(0)}(\mathrm{i}S)\,  f^{(0)}(\mathrm{i}\bar S)\,. 
\end{equation}
Subsequently we consider the function 
\begin{equation}
  \label{eq:def-F}
  F(Y) = - \frac{Y^1Y^2Y^3}{Y^0}\,,
\end{equation}
and establish that 
\begin{equation}
  \label{eq:F-asymt}
  \mathcal{F}_0(p,\phi) =  4\,\mathrm{Im}[F(Y)] \,. 
\end{equation}
The mixed free energy $\mathcal{F}_\mathrm{E}(p,\phi)$ of the STU
model equals,
\begin{eqnarray}
  \label{eq:mixed-free-STU} 
  \mathcal{F}_{\rm E} (p,\phi)\! &=&\!\mathcal{F}_0(p,\phi) \nonumber\\
  &&
  -\frac1\pi\left[ \ln\omega(p^0,p^1,\phi^0,\phi^1)
  +\ln\omega(p^0,p^2,\phi^0,\phi^2)
  +\ln\omega(p^0,p^3,\phi^0,\phi^3)\right] \,.\;
\end{eqnarray}

Now consider the limit where the charges $p^I$ and the $\phi^I$ are
large. The leading part in the prefactor in \eqref{eq:zres2} then 
equals
\begin{equation}
  \label{eq:prefactor}
  \mathrm{e}^{-\mu_s(p,\phi)} \equiv
  \frac{(p^2\phi^0 - p^0 \phi^2)
    (p^3\phi^0 - p^0 \phi^3) }  
  {((\phi^0)^2 + (p^0)^2)(\phi^0p^1-\phi^1p^0)} = \frac{(T+\bar
    T)(U+\bar U)}{2\,(S+\bar S)} \,,
\end{equation}
where we note the expression for the K\"ahler potential
$\mathcal{K}$ 
\begin{equation}
  \label{eq:Kahlerpot}
  \mathcal{K} = -\ln[(S+\bar S)(T+\bar T)(U+\bar U)]  =
    - \ln\left[\frac{\mathrm{i}(\bar Y^I F_I - Y^I \bar F_I)} {\vert
    Y^0|^2}\right]  \,,
\end{equation}
where $F_I=\partial F/\partial Y^I$. 

\section{The mixed partition function for the STU model}
\label{sec:mixed-part-funct-STU}
\setcounter{equation}{0}

In this section, we evaluate the full mixed partition function
$Z_\mathrm{STU}(p,\phi)$ for the STU model. In order to make use of the results
obtained in the previous section, we write $Z_\mathrm{STU}(p,\phi)$ as
\begin{eqnarray}
  \label{eq:stu-conv}
  Z_\mathrm{STU}(p, \phi) &=& \sum_{\{q\}} d_\mathrm{STU}(q,p) \, {\rm
    e }^{\pi q_I \, \phi^I} \nonumber\\[.1ex] 
  &=& \sum_{\{q,q',q''\}} \delta_{q,q'} \, \delta_{q',q''} \,
  d_0(K_s,L_s,M_s) \, d_0(K_t,L_t,M_t)\, d_0(K_u,L_u,M_u) \nonumber\\
  && \hspace{1.2cm} \times {\rm e}^{ \frac{\pi}{3} \left[
      \left(q_0 + q'_0 + q''_0 \right) \phi^0 +
      \left(q_1 + q'_1 + q''_1 \right) \phi^1 +
      \left(q_2 + q'_2 + q''_2 \right) \phi^2 +  \left(q_3 + q'_3 +
    q''_3 \right) 
      \phi^3 \right] } \;, \nonumber\\
  &&{~}
\end{eqnarray}
with $(K_s, L_s, M_s)$ given by \eqref{eq:KLM-pq}, and where $(K_t,
L_t, M_t)$ and $(K_u, L_u, M_u)$ follow by triality, except that at
the same time we change the charges $q$ to $q'$ and $q''$,
respectively, 
\begin{eqnarray}
  \label{eq:K-tu-prime} 
  3\,K_t &=& -2 (p^0 q'_2  + p^1 p^3) \;, \nonumber\\
  3\,L_t &=& 2 (q'_0 p^2 - q'_1 q'_3 ) \;, \nonumber\\
  3\,M_t &=&  q'_0 p^0 - q'_2 p^2 + q'_1 p^1 + q'_3 p^3 
  \;,\nonumber\\[2mm]
  3\,K_u &=& -2 (p^0 q''_3  + p^1 p^2) \;, \nonumber\\
  3\,L_u &=& 2 (q''_0 p^3 - q''_1 q''_2 ) \;, \nonumber\\
  3\, M_u &=& q''_0 p^0 - q''_3 p^3 + q''_2 p^2 + q''_1 p^1 \;.
\end{eqnarray}
We will be assuming that all charges $p^I$ are nonzero and positive. 

The insertion of the Kronecker deltas leads to three copies (one for
each of the three sectors S, T and U) of the mixed partition function
computed in section \ref{sec:prot-mixed-part}. These copies,
$Z_s(p,\phi_s)$, $Z_t(p,\phi_t)$ and $Z_u(p,\phi_u)$, are related to
the each other by triality. Using the representation for the delta
symbol (with integers $n,m$),
\begin{equation}
  \label{eq:5}
  \delta_{mn} = \int_0^{1} \mathrm{d}\theta\;
  \mathrm{e}^{2\mathrm{i} \pi (m-n) \theta} \,,
\end{equation}
we rewrite \eqref{eq:stu-conv} as follows,
\begin{equation}
  \label{eq:full-mixed-pf}
  Z_\mathrm{STU}(p, \phi) =  \int_0^{1}
  \mathrm{d}^4\theta\, 
  \mathrm{d}^4\varphi \; Z_s(p,\phi_s) \, Z_t(p,\phi_t)
  \,Z_u(p,\phi_u) \;, 
\end{equation}
where
\begin{eqnarray}
  \label{eq:shifts-phi-til}
  \begin{array}{rcl}
    \phi_s{}^0 &=&\phi^0 + 6 \mathrm{i}\lambda\,\theta^0 \;, \\
    \phi_t{}^0 &=& \phi^0 + 6 \mathrm{i}\lambda\,(\varphi^0-\theta^0)
    \;,\\
  \phi_u{}^0 &=& \phi^0 - 6\mathrm{i}\lambda \,\varphi^0 \;,
    \end{array}
\qquad
  \begin{array}{rcl}
    \phi_s{}^{1,2,3} &=&\phi^{1,2,3} + 6
    \mathrm{i}\lambda^{-1}\,\theta^{1,2,3} \;, \\ 
    \phi_t{}^{1,2,3} &=& \phi^{1,2,3} + 6
    \mathrm{i}\lambda^{-1}\,(\varphi^{1,2,3}-\theta^{1,2,3}) 
    \;,\\
  \phi_u{}^{1,2,3} &=& \phi^{1,2,3} - 6\mathrm{i}\lambda^{-1}
    \,\varphi^{1,2,3} \;. 
    \end{array}
\end{eqnarray}
Observe that
\begin{equation}
  \label{eq:sum-phis}
  \phi_s{}^I + \phi_t{}^I + \phi_u{}^I = 3\,\phi^I \;.
\end{equation}

We remind the reader that each of the factors $Z_s$, $Z_t$ and $Z_u$
is invariant under the shifts $\phi^{1,2,3} \rightarrow \phi^{1,2,3} +
6 \mathrm{i} \lambda^{-1}$ and $\phi^0 \to \phi^0+
6\mathrm{i}\lambda$, by virtue of the (finite or infinite) explicit
sums contained in these factors. Note that the infinite shift sums
occur for $\phi_s{}^{2,3}$, $\phi_t{}^{1,3}$, and $\phi_u{}^{1,2}$,
while the remaining shift sums cover a finite range. As it turns out
most of these sums can be generated by extending the integrals over
$\theta^I$ and $\varphi^I$ from the interval $[0,1]$ to a larger
interval. To explain this, consider the integration over $\varphi^{2}$
and $\theta^2$.  The factors $Z_s$ and $Z_u$ contain both an infinite
sum of shifts of $\phi^2$, whereas $Z_t$ contains a finite sum of such
shifts. The two infinite sums are thus included by extending the range
of integration of $\varphi^2$ and $\theta^2$ from $[0,1]$ to
$[-\infty, \infty]$.  In this way we are left with one finite sum over
shifts of $\phi^2$ (whose range is determined by the value of $p^0$)
and two integrals ranging over $\varphi^2,\theta^2\in[-\infty,
\infty]$.  The same procedure applies to the integration over
$\varphi^{1,3}$ and $\theta^{1,3}$.  Concerning the integration over
$\varphi^0$ and $\theta^0$, the situation is slightly different,
because each of the three factors $Z_s, Z_t$ and $Z_u$ involves a
finite sum of shifts of $\phi^0$, and each is invariant under $\phi^0
\rightarrow \phi^0 + 12\mathrm{i} p^a$ with $a=1,2,3$, respectively.
This implies that the first of the three finite sums can be used to
extend the range of integration of $\theta^0$ from $[0,1]$ to $[0, 2
p^1\lambda^{-1}]$, the second sum can be used to extend the range of
integration of $\varphi^0$ to $[0, 2 p^2\lambda^{-1}]$, while the
third sum is kept untouched. Here we may be overcounting slighly as we
explained in the text below \eqref{eq:mpfpk}, depending on the choice
for $\lambda$, but this does not present a problem of principle. Below
we will evaluate the resulting expression for large charges and large
potentials, in which case one extends all the ranges of integration to
the infinite interval $[-\infty,\infty]$ and sums over all the shifts
$\phi^{1,2,3} \rightarrow \phi^{1,2,3} + 6 \mathrm{i} \lambda^{-1}$
and $\phi^0 \to \phi^0+ 6\mathrm{i}\lambda$ at the end. Hence we
consider the following integral,
\begin{equation}
  \label{eq:full-mixed-pf-int}
  Z_\mathrm{STU}(p, \phi) =  \sum_{\phi^I{\rm-shifts}}
\int_{-\infty}^{\infty}
  \mathrm{d}^4\theta\,
  \mathrm{d}^4\varphi \; Z_s(p,\phi_s) \, Z_t(p,\phi_t)
  \,Z_u(p,\phi_u) \;,
\end{equation}
where $Z_s$, $Z_t$ and $Z_u$ follow from \eqref{eq:zres2}, but without
the explicit sum over the imaginary shifts, which have now been
incorporated in the sum over the $\phi^I$-shifts and in the extended
$\varphi^I$- and $\theta^I$-integration domains. We will evaluate this
integral in saddle-point approximation. Before doing so we consider
the integrand in somewhat more detail,
\begin{eqnarray}
  \label{eq:ZZZ}
  &&Z_s(p,\phi_s) \, Z_t(p,\phi_t)
  \,Z_u(p,\phi_u)\approx  \nonumber\\
  && \quad=  \exp\left\{\tfrac13\pi [\mathcal{F}_0(p,\phi_s) +
  \mathcal{F}_0(p,\phi_t) +\mathcal{F}_0(p,\phi_u)]\right.\nonumber\\
  &&{}\hspace{15mm}
  - [\ln\omega(p^0,p^1,\phi^0_s,\phi^1_s)
  +\ln\omega(p^0,p^2,\phi^0_t,\phi^2_t)
  +\ln\omega(p^0,p^3,\phi^0_u,\phi^3_u) ]\nonumber\\ 
  && \hspace{15mm} \left.
  - [\mu_s(p,\phi_s) +\mu_t(p,\phi_t) +\mu_u(p,\phi_u) ] \right\}\,,
\end{eqnarray}
where the $\mathcal{F}_0$ was defined in
\eqref{eq:mixed-free-energy-stu}, whereas the expressions for $\omega$
and $\mu$ follow from the ones given in \eqref{eq:Delta-divisor} and
\eqref{eq:prefactor} by triality. Here we suppressed the terms in
\eqref{eq:zres2} that vanish in the limit of large charges $p^I$ and
large potentials $\phi^I$. In that same limit, the contributions
contained in $\omega$ and $\mu$ are subleading relative to those
contained in $\mathcal{F}_0$ and can therefore be ignored when
evaluating \eqref{eq:full-mixed-pf-int} in saddle-point approximation.
We therefore expand $\mathcal{F}_0(p,\phi_s) + \mathcal{F}_0(p,\phi_t)
+\mathcal{F}_0(p,\phi_u)$ in powers of $\theta^I$ and $\varphi^I$. The
terms linear in $\theta^I$ and $\varphi^I$ all cancel out by virtue of
\eqref{eq:sum-phis}, so that we have a saddle point at $\theta^I =
\varphi^I = 0$. The term quadratic in $\theta^I$ and $\varphi^I$ is
homogeneous of zeroth degree in $(p^I,\phi^I)$, whereas higher powers
in $\theta^I$ and $\varphi^I$ have coefficients that are homogeneous
of negative degree. This indicates that possible other saddle points
will be exponentially suppressed. Retaining only the terms quadratic
in $\theta^I$ and $\varphi^I$ one may perform the corresponding
eight-dimensional gaussian integral, which turns out to be equal
(possibly up to a multiplicative constant) to $\exp[2\,\mathcal{K}]$,
where ${\cal K }$ is given by \eqref{eq:Kahlerpot}. The calculation
leading to this result is rather non-trivial. An easier exercise is to
derive this result in the special case of $p^0=0$.

Combining this result with the terms independent of $\theta^I$ and
$\varphi^I$ thus leads to the result,
\begin{equation}
  \label{eq:partition-conj}
  Z_\mathrm{STU}(p,\phi) \approx \sum_{\phi-{\rm shifts}}
  \, {\rm e}^{ \pi \, \mathcal{F}_{\rm E} (p,\phi) + {\cal K} } \,,
\end{equation}
up to an overall numerical constant. Here the mixed free energy,
$\mathcal{F}_\mathrm{E}(p,\phi)$, was defined in
\eqref{eq:mixed-free-STU}, and we used that $\exp[-
\mu_s(p,\phi) -\mu_t(p,\phi) -\mu_u(p,\phi)] = \frac18 \exp[-{\cal K}]
$. The multiplicative factor $\exp[{\cal K}]$ is in precise agreement
with the one conjectured in \cite{Cardoso:2008fr} on the basis of
semiclassical arguments for a class of $N=2$ theories which includes
the STU model.

\section{Discussion and conclusions}
\label{sec:conclusion}
\setcounter{equation}{0}
The result obtained in the previous section demonstrates that the
proposal of \cite{David:2007tq} for the dyonic degeneracies of the STU
model leads to the mixed partition function with a prefactor
that agrees with the prediction of \cite{Cardoso:2008fr}. The result
was obtained in the case that all charges $p^I$ are non-zero and
positive, in the limit of large charges and large potentials $\phi^I$.
The charges were only taken positive to simplify the formulae, and we
expect that there exists a similar result for $p^0=0$. In the latter
case, an alternative, but rather similar, calculation seems possible
provided that $p^1,p^2,p^3\not=0$.  Based on previous experience
\cite{Shih:2005he,LopesCardoso:2006bg}, we expect an analogous result.

The agreement that we have established here lends further support to
the approaches taken in \cite{David:2007tq} and \cite{Cardoso:2008fr},
and goes beyond the fact that the leading and subleading contributions
to the entropy are in agreement (up to certain subtleties that we will
again discuss below). The two approaches are based on entirely
different considerations.  Unlike in $N=4$ models, we were forced to
rely on a saddle-point approximation of the integral
\eqref{eq:full-mixed-pf-int} at the end of the calculation, but the
major part of the calculation does not depend on that. Therefore the
result could a priori have been different. In fact there are other
predictions in the literature \cite{Denef:2007vg} for the prefactor in
\eqref{eq:partition-conj}, derived in a different regime. For a
variety of reasons it seems unlikely that the present calculation can
shed some light on these different results. Some of these reasons are
discussed below.

As was stressed in \cite{Cardoso:2008fr}, there is a distinct
difference between the dyonic degeneracies for the various $N=4$
models proposed earlier and the expression for the dyonic degeneracies
in the STU model, which was already exhibited in \cite{David:2007tq}.
The remarkable feature of the $N=4$ models is that the saddle-point
equations for the leading and subleading terms (c.f.
\eqref{eq:value-dk-saddle}), which determine the entropy of large
black holes from the microscopic degeneracies, coincide with the
attractor equations of supergravity
\cite{Cardoso:2004xf,David:2006yn}. This feature might be due to the
high degree of symmetry in $N=4$ models. For the STU model this
relationship does not hold, although the statistical and the
macroscopic entropy still agree to this order. Though this difference
in behaviour of the dyonic degeneracy formula for the STU model from
that of the $N=4$ models does not, perhaps, indicate any fundamental
inconsistency, it warrants at least a closer study of the next
subleading correction to the entropy.

Another remarkable feature of the dyonic degeneracy formula for $N=4$
models is that its form remains the same across walls of marginal
stability. The dependence of the degeneracies on the asymptotic moduli
is encoded in the choice of the integration contour used for
extracting the degeneracies from \eqref{eq:dphik}
\cite{Sen:2007vb,Sen:2007pg,Cheng:2007ch}.  When the asymptotic moduli
cross walls of marginal stability, the dyon can decay into a pair of
$1/2$-BPS states.  Let us focus on a wall of marginal stability at
which a $1/4$-BPS dyon decays into a pair of purely electric and
purely magnetic $1/2$-BPS states,
\begin{equation}
  \label{eq:decay}
  (Q,P) \rightarrow (Q,0) + (0, P)\,.
\end{equation}
Then, by general arguments \cite{Denef:2000nb,Denef:2007vg}, the
degeneracy of $1/4$-BPS states jumps across such a wall and the change
is given by
\begin{equation}
   \label{eq:wallcross}
   d_>(Q, P) - d_<(Q,P)  = (Q\cdot  P)\, (-1)^{(Q\cdot P)+ 1}\,
   d_\mathrm{el}(Q) \, d_\mathrm{mag}(P) \,, 
\end{equation}
where $d_>(Q,P)$ and $d_<(Q,P)$ refer to the degeneracies of $1/4$-BPS
states across the wall, and $d_\mathrm{el}(Q)$ and $d_\mathrm{mag}(P)$
refer to the degeneracy of the purely electric and purely magnetic
$1/2$-BPS states.  This wall crossing formula is obeyed by the $N=4$
dyon degeneracy formula, because the modular form factorizes across
the divisor $\upsilon=0$ as
\begin{equation}
  \label{eq:doublepole}
  \Phi_k(\rho, \sigma,\upsilon)   \sim 4\pi^2\,
  \upsilon^2 \,g^{(k)}_\mathrm{mag}(\rho)\,
  g^{(k)}_\mathrm{el}(\sigma) \,, 
\end{equation}
where $g^{(k)}_\mathrm{el}(\sigma)$ denotes the partition function for
purely electric states and $g^{(k)}_{mag}(\rho)$ denotes the partition
function for purely magnetic states. Then the jump in
(\ref{eq:wallcross}) arises due to the contribution of the double pole
at $\upsilon\sim 0$. This feature of the $N=4$ degeneracy formulae
ensures that the function retains the same form across a wall of
marginal stability and that the degeneracies can just be extracted by an
appropriate choice of the integration contour.
 
Let us examine whether the above feature is present in the partition
function of dyons in the STU model. The STU model also admits a wall
of marginal stability at which the dyon decays according to
\eqref{eq:decay}, and we may consider whether the corresponding
automorphic form admits a similar factorization as in
\eqref{eq:doublepole}. Since the partition function is a product of
three modular forms $\Phi_0(\rho,\sigma,\upsilon)$, there are three
divisors, $\upsilon_s =0$, $\upsilon_t=0$ and $\upsilon_u=0$. At, say,
the divisor $\upsilon_s=0 $ and $\upsilon_t, \upsilon_u\neq 0$, the
degeneracy formula factorizes as (see \cite{David:2007tq} for the
properties of $\Phi_0$)
 \begin{eqnarray}
   \label{eq:em-divisor} 
 && \Phi_0(\rho_s, \sigma_s,\upsilon_s)\;
  \Phi_0(\rho_t,\sigma_t, \upsilon_t)\;
  \Phi_0(\rho_u,\sigma_u,\upsilon_u)   \\ \nonumber
 &&\qquad 
 \sim 4\pi^2\, \upsilon_s{}^2 \;\frac{\eta^8(2\rho_s)}{\eta^4(\rho_s)}\;
  \frac{\eta^8(\sigma_s/2)}{\eta^4(\sigma_s)} \;
 \Phi_0(\rho_t,\sigma_t,
  \upsilon_t)\;\Phi_0(\rho_u,\sigma_u,\upsilon_u) \;. 
\end{eqnarray}
The contribution of this double pole to the degeneracy is of the form
\begin{equation}
  \label{eq:stuwall}
  M_s\, ( -1)^{M_s +1} \; d_1(K_s)\; d_2 (L_s)\; I_0(K_t,
  L_t, M_t) \;I_0(K_u, L_u, M_u) \;, 
\end{equation}
where
\begin{equation}
  \label{eq:mag-el-degen} 
  d_1(K_s) = \oint 
  \mathrm{d}\rho \;\frac{\mathrm{e}^{\mathrm{i}\pi
  K_s\rho}}{\eta^8(2\rho)\, \eta^{-4}(\rho)}  \,, \qquad
  d_2(L_s) = \oint \mathrm{d}\sigma
  \frac{\mathrm{e}^{\mathrm{i}\pi L_s\sigma} } 
  {\eta^8(\sigma/2)\, \eta^{-4}(\sigma)}\;. 
\end{equation}
Certainly (\ref{eq:stuwall}) does not obey the wall crossing formula
(\ref{eq:wallcross}). The same conclusion holds at the other divisors
$\upsilon_t=0$ or $\upsilon_u=0$, or combinations thereof. This
suggests that the degeneracy formula \eqref{eq:deg-STU} is valid only
in the region of asymptotic moduli where the singlecentered black hole
is stable.  Restricting the domain of validity of the partition
function to such a region avoids the entropy enigma, because the
multicentered solutions found by \cite{Denef:2007vg}, which dominate
the entropy, are not stable in that case. It will be interesting to
study this region by carefully considering the walls of marginal
stability for the STU model. Across walls of marginal stability, the
partition function should be modified in such a way that the wall
crossing formula holds.  Therefore such a study can perhaps indicate
how the degeneracy formula can be extended to other domains.

\subsection*{Acknowledgements}
We acknowledge helpful discussions with Thomas Mohaupt and Ashoke Sen.
G.L.C. and B.d.W. thank the CERN Theory Black Hole Institute, where
part of this work was carried out, for hospitality. This work is
partly supported by EU contracts MRTN-CT-2004-005104 and
MRTN-CT-2004-512194 and by NWO grant 047017015.

\providecommand{\href}[2]{#2}
\begingroup\raggedright\endgroup
\end{document}